\documentclass[letter]{aa} 
\usepackage{ae,lmodern} 
\usepackage{txfonts}
\usepackage{ulem}
\usepackage[]{xcolor}
\usepackage[colorlinks,allcolors=blue]{hyperref}

\begin{document} 
\title{FAUST XXX:}\subtitle{Dust enhancement in the young binary L1551 IRS~5}
\titlerunning{Dust enhancement in the young binary L1551 IRS~5}
   \author{N. Cuello\inst{1}
\and E. Bianchi\inst{2,3,1}
\and F. M\'enard\inst{1}
\and L. Loinard\inst{4}
\and R. Hernández Garnica\inst{4}
\and A. Durán\inst{4}
\and C. Ceccarelli\inst{1} \and
\newline M.J. Maureira\inst{5} 
\and C.J. Chandler\inst{6}
\and C. Codella\inst{2,1}
\and N. Sakai\inst{7}
\and L. Podio\inst{2}
\and G. Sabatini\inst{2}
\and L. Chahine\inst{1} \and
\newline M. de Simone\inst{8}
\and D. Fedele\inst{2}
\and D. Johnstone\inst{9,10}
\and T. Hanawa\inst{11}
\and I. Jiménez-Serra\inst{12}
\and S. Yamamoto\inst{13}
}
\authorrunning{Cuello et al.}
\institute{Univ. Grenoble Alpes, CNRS, IPAG, 38000 Grenoble, France
\and
INAF, Osservatorio Astrofisico di Arcetri, Largo E. Fermi 5, I-50125, Firenze, Italy
\and
Excellence Cluster ORIGINS, Boltzmannstraße 2, D-85748, Garching bei München, Germany
\and
Instituto de Radioastronomía y Astrofísica, Universidad Nacional Autónoma de México, A.P. 3-72 (Xangari), 8701, Morelia, Mexico
\and
Max-Planck-Institut für extraterrestrische Physik (MPE),
Gießenbachstr. 1, D-85741 Garching, Germany,
\and
National Radio Astronomy Observatory, PO Box O, Socorro, NM 87801,
USA
\and
RIKEN Cluster for Pioneering Research, 2-1, Hirosawa, Wako-shi, Saitama 351-0198, Japan
\and
ESO, Karl Schwarzchild Srt. 2, 85478 Garching bei München, Germany
\and
NRC Herzberg Astronomy and Astrophysics, 5071 West Saanich Road, Victoria, BC, V9E 2E7, Canada
\and
Department of Physics and Astronomy, University of Victoria, Victoria, BC, V8P 5C2, Canada
\and
Center for Frontier Science, Chiba University, 1-33 Yayoi-cho, Inage-ku, Chiba 263-8522, Japan
\and
Centro de Astrobiología (CSIC/INTA), Ctra. de Torrejón a Ajalvir km 4, 28806, Torrejón de Ardoz, Spain
\and
SOKENDAI, Shonan Village, Hayama, Kanagawa 240-0193, Japan
}
   \date{Received 28/11/2025; Accepted 10/12/2025}

\abstract{Young binary stars with discs provide unique laboratories to study the earliest stages of planet formation in star-forming environments. The detection of substructure in discs around Class I protostars challenges current models of disc evolution, suggesting that planets may form earlier than previously expected ($<1$~Myr). In the context of the ALMA Large Program FAUST, we present observations of the circumbinary disc (CBD) around the young binary system L1551 IRS 5. The CBD exhibits two prominent over-densities in the continuum emission at the edge of the cavity, with the Northern over-density being about 20\% brighter than the Southern one. By analysing the disc morphology and kinematics of L1551 IRS~5, we delineate dynamical constraints on the binary’s orbital parameters. Additionally, we present 3D hydrodynamical models of the CBD to predict both the dust and the gas surface densities. Then, we compare the resulting synthetic observations with ALMA observations of the continuum emission at 1.3~mm and the C$^{18}$O line emission. Our analysis suggests that the density enhancements observed with ALMA in L1551 IRS~5 can be caused by interactions between the binary stars and the CBD, leading to dust concentration within the disc. We conclude that the observed over-density corresponds to a location where could potentially grow under favourable conditions.}
   \maketitle
\section{Introduction}
\label{sec:intro}

The process of star formation is intimately linked with the emergence of binary star systems with discs \citep{Duchene&Kraus2013}. The collapse of a molecular cloud under its own gravity triggers the formation of protostars with surrounding discs made of gas and dust \citep{Bate2018, Lebreuilly+2021, Kuruwita+2023}. This process naturally leads to a high fraction of young multiple stellar systems, where stars orbit around each other \citep{Reipurth+2014, Offner+2023}. The complex interplay between gravity and radiation explains the large variety of structures observed in discs around young stars \citep{Tobin+2016, Tobin+2022}. In this context, the study of binaries with discs -- both through observations and simulations --- is crucial to reveal where, how, and when planets could form around young stars.

The advent of new generation telescopes, such as the Atacama Large Millimeter/submillimeter Array (ALMA), enabled the detection of asymmetric millimetre continuum emission around several Myr-old stellar systems. The systems IRS~48 \citep{vanderMarel+2013, Calcino+2019} IRAS04158+2805 \citep{Villenave+2020, Ragusa+2021} and HD~142527 \citep{Casassus+2015, Price+2018} constitute prototypical examples of horseshoe-shaped asymmetries in circumbinary discs (CBDs). These over-densities of millimetric dust can be explained by the periodic forcing of an inner binary. This mechanism translates into a localised pile-up of material at the inner edge of the disc \citep[e.g.,][]{Ragusa+2017, Thun+2017, Poblete+2019, Hirsh+2020}. However, not all azimuthal asymmetries correspond to long-lived dust traps: some are transient gas over-densities at the apocentre of eccentric discs, while others reflect short-lived clumps produced by dynamical perturbations \citep{Ragusa+2020}. Hence, it is important to distinguish between persistent pressure maxima (causing dust traffic jams) and transient circumbinary structures.

These binary-induced over-densities can either orbit at the local Keplerian frequency or precess on secular timescales of a few hundred binary orbital periods \citep{Ragusa+2020, Munoz&Lithwick2020}. The former are able to efficiently concentrate the dust density around the binary. Therefore, the presence of such dust over-densities in CBDs should have important consequences for dust growth --- and eventually planetesimal formation. Still, until now, most studies have focused on Class II stars with such asymmetries \citep{Bae+2023}.

Moreover, images of discs around the younger (Class I, <~1~Myr) protostars HL~Tau \citep{HLTAU+2015}, GY 91 \citep{Sheehan+2018}, and IRS~63 \citep{Segura-Cox+2020} have revealed conspicuous structures similar to those found in more evolved (Class II) T~Tauri and Herbig~AeBe stars \citep[e.g.,][]{Andrews+2018, Long+2018}. These discoveries have unambiguously brought along the idea that planets --- possibly responsible for the observed gaps and rings --- might actually form earlier than expected. A further indication is provided by the estimate mass of solids in Class II discs, which is insufficient to explain the observed exoplanets distribution \citep{Manara+2018, Tychoniec+2020}. Observations of substructures in young Class 0/I discs are so far limited because of the challenge of high dust opacity at mm-wavelengths \citep{Ohashi+2023}. Yet, young protostars (often multiple) constitute unique laboratories to test the early stages of planet formation. 

In this study, we focus on the circumbinary disc (CBD) around the Class I protostar L1551 IRS~5 \citep{Adams+1987, Looney+1997}. This system is located in Taurus \citep{Strom+1976} at a distance of 146.4 $\pm$ 0.5 pc \citep{Galli+2018, Galli+2019} and has a luminosity of $L_{\rm bol} = 26.6 \, L_\odot$ \citep{Green+2013}. L1551 IRS~5 is surrounded by a large rotating and infalling envelope, whose motion was measured through lines from several species \citep[e.g.,][]{ White+2006}. Its binary nature was first revealed by \cite{Bieging+1985}, where the northern and southern protostars have masses of $0.8\,M_\odot$ and $0.3\,M_\odot$, respectively \citep{Looney+1997, Liseau+2005, Lim+2016}. In addition, \citet{Rodriguez+2003b} reported two parallel jets in this system. The binary is surrounded by a CBD which size and mass are respectively around $140$~au and $0.02-0.03\,M_\odot$ \citep{Looney+1997, Cruz-SaenzdeMiera+2019}. The CBD inclination and position angle are estimated using a 2D Gaussian fitting modelling and are \textit{i}= 60$\degr$ $\pm$ 2$\degr$  and PA=161$\degr$ $\pm$ 2$\degr$, respectively \citep{Cruz-SaenzdeMiera+2019}. Based on ALMA observations from the {\sc faust} (Fifty AU STudy of the chemistry in the disc/envelope system of solar-like protostars\footnote{http://faust-alma.riken.jp}) Large Programme \citep{Codella+2021}, \cite{Bianchi+2020} reported the discovery of a hot corino around the Northern (primary) star and possibly around the Southern (secondary) star. More recently, additional ALMA observations indicate the presence of two compact dusty discs around each stellar component \citep{Maureira+2025}. The orbit of the L1551 IRS~5 binary (with an orbital period of about 260~yr, first constrained by \citealt{Lim+2016}) was also refined by combining nearly forty years of VLA and ALMA astrometric measurements by \cite{Hernandez+2024} who obtained a binary semi-major axis of $\sim44$~au. The mass of the system is constrained to be $0.96 \pm 0.17$ $M_\odot$, and the eccentricity is found to be slightly below $0.3$. Previous studies have suggested that the binary-disc interaction is responsible for the asymmetrical disc morphology observed \citep{Takakuwa+2017, Matsumoto+2019, Takakuwa+2020}.

In this article, we present new ALMA observations of L1551 IRS~5 in Sect.~\ref{sec:observations} and tailored hydrodynamical models of the ongoing binary-disc interaction along with synthetic observations in Sect.~\ref{sec:simulations}. We discuss the implications of our results for disc evolution and planet formation around young binary stars in Sect.~\ref{sec:discussion}.

\section{FAUST observations of L1551 IRS 5}
\label{sec:observations}

L1551-IRS5 was observed by ALMA, between October 2018 and January 2019, as part of the {\sc faust} Large Program \citep[project 2018.1.01205.L; PI: S. Yamamoto,][]{Codella+2021}. The phase centre was at the coordinates $\alpha_{\rm J2000}$ = 04$^{\rm h}$ 31$^{\rm m}$ 34$\fs$14, $\delta_{\rm J2000}$ = +18$\degr$ 08$\arcmin$ 05$\farcs$10. The continuum data presented here were acquired using the most extended {\sc faust} configuration (C43-5), with baselines ranging between 15~m and 1.4~km. The C$^{18}$O (2-1) at 219.56036 GHz (E$_{\rm up}$= 16 K) was observed in Band 6, in two 12-m array configurations: the most extended with baselines ranging between 15 m and 1.4 km (C43-5), and an intermediate with baselines between 15~m and 484~m (C43-2). Observations were complemented using the 7-m Atacama Compact Array (ACA), with baselines between 8.9~m and 48.9~m. For the most extended configuration and the ACA observations, the band-pass and flux calibrator is the quasar J0423-0120, while the phase calibrator is J0510+1800. For the intermediate configuration, the band-pass and flux calibrator is J0510+1800, while the phase calibrator is J0440+1437. The spectral window covering the C$^{18}$O (2-1) transition has a spectral resolution of 122 kHz ($\sim$0.17 km s$^{-1}$). Data were calibrated using the ALMA calibration pipeline in {\sc casa} \citep{CASA-software}, and an additional calibration routine to correct for the system temperature (T$_{\rm sys}$) and spectral line data normalisation\footnote{\url{https://help.almascience.org/index.php?/\\Knowledgebase/Article/View/419}}. Finally, self-calibration was performed using line-free continuum channels. The uncertainty on the absolute flux calibration is about 10\%.

\begin{figure*}
    \centering
   \includegraphics[width=\textwidth]{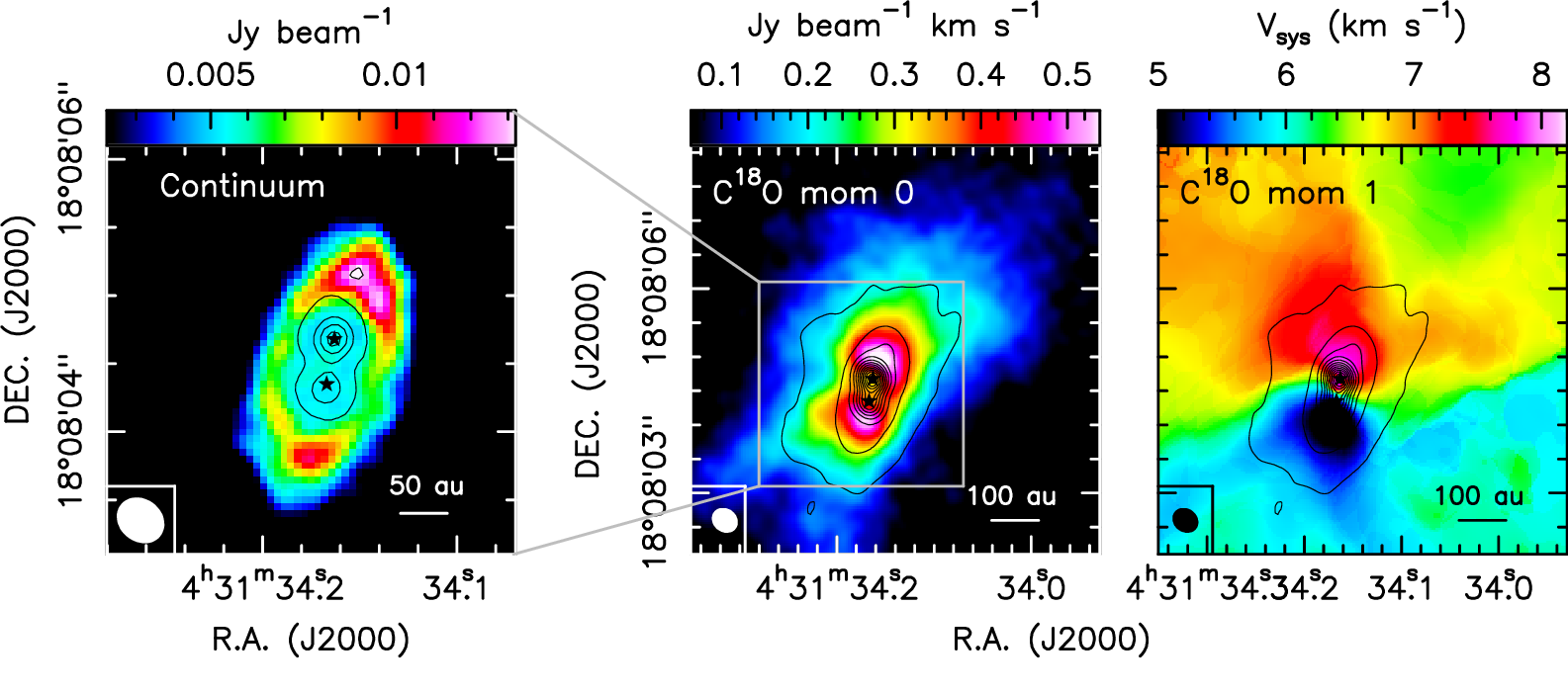}
    \caption{\textit{Left:} Continuum emission at 1.33 mm in colour scale (see Sec. \ref{sec:observations}). Contours indicate the protostellar discs, subtracted from the image during the cleaning process. First contours and steps are 30$\sigma$ (13 mJy beam$^{-1}$) and 100$\sigma$, respectively. The black stars indicate the positions of the N and S protostars.
    \textit{Middle:} Moment 0 map of the C$^{18}$O emission (integrated between -10 km s$^{-1}$ and +20 km s$^{-1}$) in colour scale superposed to 1.3 mm dust continuum emission in grey contours (from \citealt{Bianchi+2020}). First contours and steps are 10$\sigma$ (1.8 mJy beam$^{-1}$) and 100$\sigma$, respectively. 
    \textit{Right:} Moment 1 map of the C$^{18}$O emission in colour scale, 1.3 mm dust continuum emission in grey contours. The moment 1 map is generated using a threshold of 5$\sigma$ (20 mJy beam$^{-1}$). The source systemic velocity ($v_{\rm sys}$) is 6.4\,${\rm km\,s^{-1}}$ \citep{Mercimek+2022}}
    \label{fig:panel-obs}
\end{figure*}

The dust emission in L1551 IRS5 shows two compact sources and an extended ($\sim$ 1$\arcsec$, corresponding to 146 au at the source distance) circumbinary disc \citep{Bianchi+2020,Cruz-SaenzdeMiera+2019, Maureira+2025}. To highlight the emission from the circumbinary disc, the imaging was performed using super-uniform weighting to give extra weight to the longest baselines. CLEAN was then manually performed placing clean boxes around the two central sources in order to remove the emission of the two circumstellar discs as detailed in \cite{Duran+2025}. The restored beam is 0$\farcs$2 ($\sim$ 30 au) and the r.m.s. is 0.5 mJy beam$^{-1}$. The resulting image is shown in Fig. \ref{fig:panel-obs} (left panel). An over density is observed in the northern part of the circumbinary disc. The maximum emission in the over-density, corresponding to the brightest area within the CBD, is at the coordinates $\alpha_{\rm J2000}$ = 04$^{\rm h}$ 31$^{\rm m}$ 34$\fs$150, $\delta_{\rm J2000}$ = +18$\degr$ 08$\arcmin$ 05$\farcs$154. Another weaker emission peak is observed in the southern part of the circumbinary disc at the position $\alpha_{\rm J2000}$ = 04$^{\rm h}$ 31$^{\rm m}$ 34$\fs$177, $\delta_{\rm J2000}$ = +18$\degr$ 08$\arcmin$ 03$\farcs$799. A flux ratio of 1.2 (0.2) is measured between the two emission peaks, which is consistent with the measured ratio of 1.4 at the same coordinates in the continuum image using the standard cleaning \citep{Bianchi+2020}. The overall emission map shows that the dusty disc extends for more than 2$\arcsec$: the inner 2$\arcsec$ are dominated by the protostellar emission, while the external part is fainter by a factor of 5-6.

Recent multi-wavelength analysis of L1551 IRS~5 \citep{Maureira+2025} shows that the dust optical depth at 1.3\,mm along the continuum enhancements is modest ($\tau_{1.3mm}\!\sim\!0.3$–0.4), excluding optical–depth saturation as their cause; spectral-index values $\beta\!\lesssim\!1$ in the southern feature further suggest local grain growth. The C$^{18}$O(2–1) emission, expected to be optically thin at $\sim$50\,au, was imaged by combining the 12m and 7m data with a robust parameter of 0.5, yielding a $0\farcs4\times0\farcs34$ beam and 2.0\,mJy\,beam$^{-1}$ rms. In Fig.~\ref{fig:panel-obs}, the moment 0 (centre) and moment 1 (right) maps reveal gas more extended than the continuum, tracing both the CBD and the envelope. Moreover, the resolved arc-like continuum morphology and the S-shaped velocity field are not consistent with projection effects alone and are characteristic of binary–disc interactions \citep{Rosenfeld+2014, Price+2018}.

\section{Modelling of the CBD in L1551 IRS~5}
\label{sec:simulations}

\subsection{Hydrodynamical simulations}
\label{sec:phantom}

We model the binary system in L1551 IRS 5 and its circumbinary disc using 3D hydrodynamical simulations with the Smoothed Particle Hydrodynamics (SPH) code Phantom \citep{phantom}, using the one-fluid approach for gas and dust mixtures and without self-gravity. The binary is represented by two sink particles with masses  $M_1 = 0.8\,M_\odot$ and $M_2 = 0.3\,M_\odot$, and its orbital parameters are sampled to be nearly coplanar with the disc and mildly eccentric (see Appendix~\ref{sec:phantom} for further details).

\begin{figure*}
    \centering
   \includegraphics[width=0.8\textwidth]{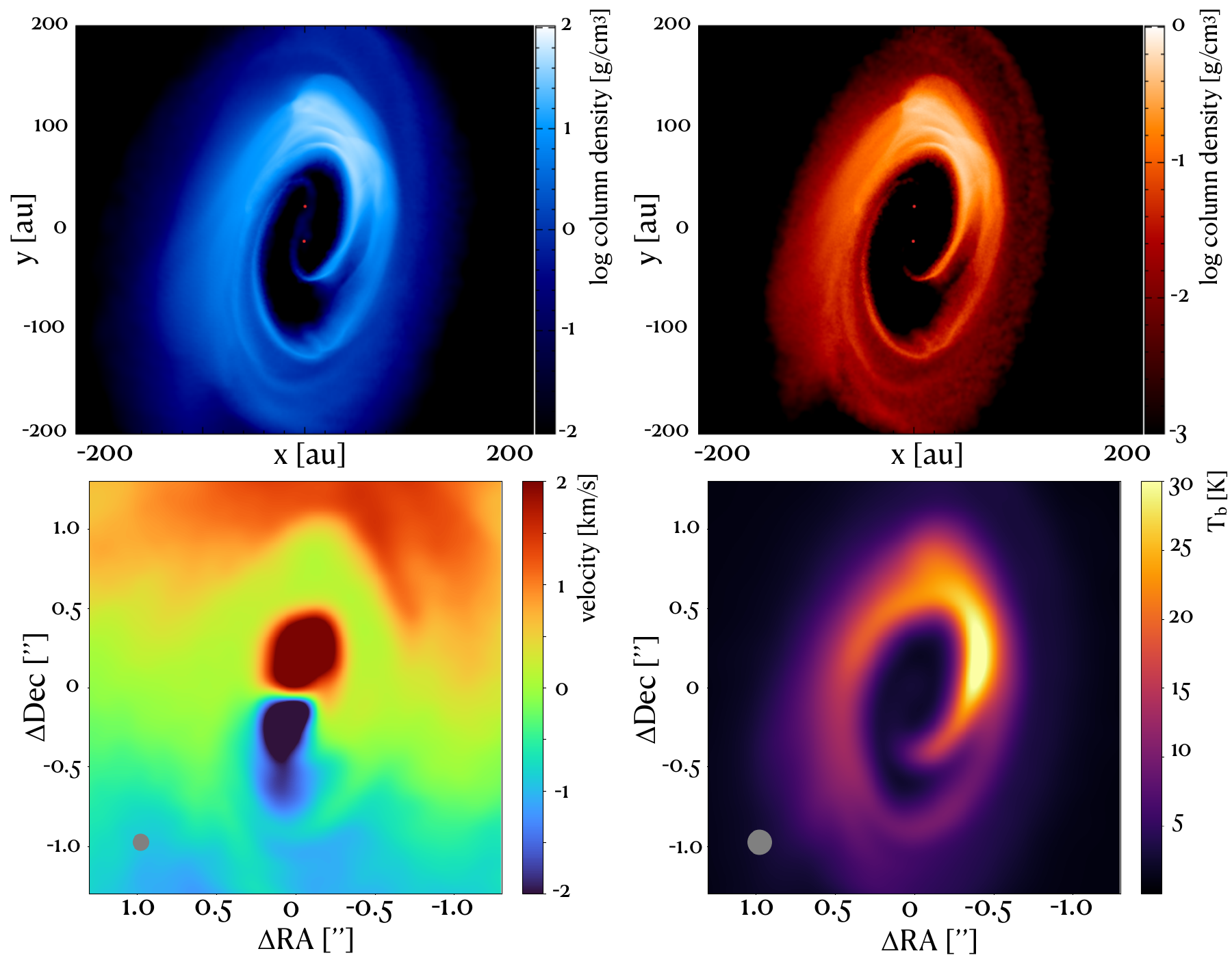}
    \caption{\textit{Top row:} {\sc Phantom} hydrodynamical models of L1551 IRS 5, gas surface density (left) and dust 100 $\mu$m-sized dust particles surface density (right). \textit{Bottom row:} {\sc mcfost} synthetic observations, C$^{18}$O moment 1 map (left) and continuum emission (right). Snapshots after 13\,100 years, which correspond to more than 50 binary orbits. All the images are shown using the same scale (top in au and bottom in arcsec).}
    \label{fig:panel-sims}
\end{figure*}

In the upper row of Fig.~\ref{fig:panel-sims}, we show the gas and dust surface density maps after 13 kyr of evolution, which corresponds to more than 50 binary orbital periods. Due to the binary interaction, the disc morphology rapidly evolves from the unperturbed state at the beginning of the simulation but reaches a steady configuration after approximately 10 binary orbital periods. From that time onwards, the disc morphology is characterised by the formation of a precessing over-density near the CBD inner edge, whose precession period is approximately equal to 1\,400~yr (roughly 5 times the binary orbital period). We note some level of variability with the binary orbital phase given that the binary is slightly eccentric and that the secondary star is closer to the CBD inner edge (compared to the primary). This leads to the formation of spiral-shaped streamers which periodically cross the orbit of the inner stars. Last, given the strong level of coupling between gas and dust particles (see Stokes numbers in Appendix~\ref{sec:phantom} and dust-to-gas plot in Fig.~\ref{fig:dgr}), we note that both surface density maps exhibit the same morphology. Hence, the horseshoe-shaped azimuthal over-density in the gas translates into a dust over-density at the exact same location. Given that the disc over-density is rotating in a Keplerian fashion and that the stars move on shorter time-scales, the proposed configuration is not unique but consistent and in excellent agreement with the current disc morphology and stars' positions. The impact of the expected CBD dynamics in L1551 IRS~5 on dust growth and stellar stellar accretion is discussed in Sect.~\ref{sec:discussion}.

\subsection{Radiative transfer models}
\label{sec:mcfost}

To compare our model with observations of L1551 IRS 5, we post-processed a subset of our simulations using the Monte Carlo radiative transfer code MCFOST \citep{Pinte+2006, Pinte+2009}. This allows us to map the hydrodynamical simulation outputs onto a radiative transfer grid without interpolation, compute thermal dust emission, and predict line emission from gas assuming local thermodynamic equilibrium (see Appendix~\ref{sec:mcfost} for details).
 
In the lower row of Fig.~\ref{fig:panel-sims}, we show the C$^{18}$O moment~1 map (left) and the continuum emission map (right) obtained with {\sc mcfost} using the methodology described above. The former is in good agreement with the rightmost panel of Fig.~\ref{fig:panel-obs}, with comparable deviations from the systemic velocity and with the same twisted pattern in the inner most region (in green). This peculiar feature is the result of the infalling material, which adds a radial velocity component within the cavity, smeared out by the beam size as described in \cite{Rosenfeld+2014}. Similar patters are routinely observed in Class II discs in which the inner disc regions are twisted and/or misaligned with respect to the outer regions. Remarkably, (sub-)stellar companions within the inner disc cavity can create such structures as in \cite{Ragusa+2021}. Last, the synthetic continuum map emission is in good agreement with the leftmost panel of Fig.~\ref{fig:panel-obs} given that we recover the prominent azimuthal over-density in the NE side of the disc. We note however that our model does not reproduce satisfactorily the over-density located on the opposite disc side, which could potentially be connected with the spiral pattern created by the inner binary in the SW disc side.

\section{Discussion and conclusions}
\label{sec:discussion}

The comparison between our model and the {\sc faust} observations of L1551 IRS5 provide evidence that the most prominent CBD features (disc size, morphology, kinematics) can be well explained by the periodic perturbation of the inner binary. In particular, the continuum over-density spatially coincides with the binary-induced gas over-density in the inner CBD edge. This supports an ongoing binary-disc interaction, in which the over-density indicates that the dust concentration is locally enhanced with respect to the rest of the CBD. However, as shown in Fig.~\ref{fig:dgr}, the dust-to-gas remains uniform --- as opposed to a dust trap where the latter increases. It is worth noting that \cite{Maureira+2025} recently reported a spectral index map for L1551 IRS~5 consistent with a mild level of grain growth within the disc. This physical mechanism is not considered in our modelling though. Considering that the disc is optically thin \citep{Takakuwa+2020, Maureira+2025}, future multi-wavelength observations will allow us to further quantify grain growth within the CBD. At any rate, our qualitative model suggests that the dust over-density is expected to be long-lived (as in \citealt{Ragusa+2020}) and that the physical conditions to efficiently concentrate dust could already be obtained during the Class 0/I phase \citep{Cridland+2022}. This could potentially lead to a disc morphology similar to IRS~48, a Class II source in which there is strong evidence of azimuthal dust trapping in the disc \citep{vanderMarel+2013, Calcino+2019}.

The method we followed in this work combines disc observations, astrometry, and hydrodynamical modelling of L1551~IRS5 in order to better constrain the inner binary orbit (mainly in terms of $a$, $e$ and $i$). In particular, among the possible orbits allowed by the astrometry in \cite{Hernandez+2024}, we selected the most consistent orbit considering the specific disc morphology obtained from observations. Although our model is designed for a specific source, this approach outlines the importance of considering constrains of different nature when modelling young multiple sources with partial time-coverage (see for instance \citealt{Duchene+2024, Toci+2024}).

In the circumbinary disc of L1551 IRS~5, we observe a distinct cavity and an azimuthal over-density at the inner edge. Accretion streamers feed material onto the binary, likely modulating stellar brightness and episodic accretion events \citep{Connelley+2014}. Unlike the models of \cite{Takakuwa+2020}, in which the cavity is only partially carved and the northern flux excess is attributed to a tightly wound southern spiral, we find that the over-density arises from direct binary-disc interaction and is not part of a large-scale spiral. In our simulations, the southern infalling spiral is more prominent than the northern one, reflecting the asymmetry observed in the system and highlighting alternative pathways for material transfer. By incorporating updated astrometric constraints, our model provides a coherent explanation for the observed disc substructures and dynamics, with fewer discrepancies compared to previous models.

In addition, the interaction between the CBD and the binary naturally channels material onto each protostars, and this enhanced accretion may be connected to the observed outflows and to the jets first identified by \cite{Itoh+2000}. Discrepancies between our hydrodynamical simulations and ALMA observations (regarding the extent of gas emission mainly), can be attributed to our simulations not incorporating the surrounding envelope, which is beyond the scope of this work. This omission results in the simulations underestimating the spatial extent of the C$^{18}$O emission, which in actual observations is influenced by contributions from both the disc and the envelope. Such insights emphasise the necessity to integrate envelope dynamics in future models to enhance the fidelity of simulated astrophysical phenomena against real observational data --- especially in young embedded systems such as L1551 IRS 5.

Thanks to upcoming morphological, chemical, and kinematical surveys that will enhance our understanding of Class 0/I protostars with unparalleled spectral and spatial detail, the field of protostellar formation is on the verge of major breakthroughs. Initiatives such as those undertaken by the {\sc faust} collaboration \citep{Codella+2021} exemplify this ongoing effort within the community. Recent studies, including the detailed characterisation of the remarkable Class I quadruple stellar system VLA 1623-2417 \citep{Ohashi+2022, Mercimek+2023, Codella+2024, Radley+2025}, underline the intricate dynamics and early evolutionary stages of these young systems. Insights obtained from such investigations are vital for unravelling the complex processes governing the formation and early development of protostellar discs. This in turn will be crucial for elucidating how dust accumulates within these discs and the subsequent emergence of embedded planets, offering a clearer picture of planetary formation around young stars.

\begin{acknowledgements}
We thank Antoine Alaguero, Daniel J. Price, and Christophe Pinte for scientific discussions and technical support. This research has made use of the NASA Astrophysics Data System. This project has received funding from the European Research Council (ERC) under the European Union Horizon Europe programme (grant agreements No. 101042275, project Stellar-MADE; No. 101053020, project Dust2Planets). NC acknowledges support from the European Union's Horizon 2020 research and innovation programme under the Marie Sk\l{}odowska-Curie grant agreement No 896319. L.L. acknowledges the support of DGAPA PAPIIT grants IN108324 and IN112820 and CONACyT-CF grant 263356. EB acknowledges support from the Deutsche Forschungsgemeinschaft (DFG, German Research Foundation) under Germany´s Excellence Strategy – EXC 2094 – 390783311. 
LP, ClCo, EB, GS acknowledge the PRIN-MUR 2020 BEYOND-2p (“Astrochemistry beyond the second period elements”, Prot. 2020AFB3FX), the project ASI-Astrobiologia 2023 MIGLIORA (Modeling Chemical Complexity, F83C23000800005), the INAF-GO 2023 fundings PROTO-SKA (Exploiting ALMA data to study planet forming discs: preparing the advent of SKA, C13C23000770005), the INAF Mini-Grant 2022 “Chemical Origins” (PI: L. Podio), INAF-Minigrant 2023 TRIESTE (“TRacing the chemIcal hEritage of our originS: from proTostars to planEts”; PI: G. Sabatini), the National Recovery and Resilience Plan (NRRP), Mission 4, Component 2, Investment 1.1, Call for tender No. 104 published on 2.2.2022 by the Italian Ministry of University and Research (MUR), funded by the European Union – NextGenerationEU– Project Title 2022JC2Y93 ChemicalOrigins: linking the fossil composition of the Solar System with the chemistry of protoplanetary discs – CUP J53D23001600006 - Grant Assignment Decree No. 962 adopted on 30.06.2023 by the Italian Ministry of Ministry of University and Research (MUR).
I.J-.S acknowledges funding from grant PID2022-136814NB-I00 funded by the Spanish Ministry of Science, Innovation and Universities/State Agency of Research MICIU/AEI/ 10.13039/501100011033 and by "ERDF/EU".
This Letter makes use of the following ALMA data: ADS/JAO.ALMA\#2018.1.01205.L (PI: S. Yamamoto). ALMA is a partnership of the ESO (representing its member states), the NSF (USA) and NINS (Japan), together with the NRC (Canada) and the NSC and ASIAA (Taiwan), in cooperation with the Republic of Chile. The Joint ALMA Observatory is operated by the ESO, the AUI/NRAO, and the NAOJ.
\end{acknowledgements}

\bibliographystyle{aa}
\bibliography{L1551.bib}

\begin{appendix} 
\section{Supplementary material}
\label{sec:supplementary}

\subsection{Hydrodynamical modelling}
\label{sec:phantom}

We model the binary in L1551 IRS~5 and its circumbinary disc through 3D hydrodynamical simulations performed with the Smoothed Particle Hydrodynamics (SPH) code {\sc Phantom} \citep{phantom}. We use SPH artificial viscosity to model angular momentum transport within the CBD, adopting $\alpha_{\rm AV}=0.25$ and $\beta_{\rm AV}=2.0$. This setup corresponds to an effective $\alpha_{\rm SS} \simeq 3-5 \times10^{-3}$ \citep{LodatoPrice2010}. The binary is modelled using two sink particles of mass $M_1=0.8\,M_\odot$ and $M_2=0.3\,M_\odot$ (see observational constraints in Sect.~\ref{sec:intro}), which corresponds to a mass ratio $q=M_2/M_1=0.375$ and a binary total mass $M_{\rm b}=1.1\,M_{\odot}$. We remove SPH particles from the simulation whenever they cross the sink radius ($R_{\rm sink}=5$~au) of one of the sink particles, as in \cite{Bate+1995}. Sink particles exert gravity on each other and are free to interact with the material from the circumbinary disc.

In order to define the binary orbit, we ran a Monte Carlo calculations using {\sc imorbel}\footnote{https://github.com/drgmk/imorbel} \citep{Pearce+2015} to sample the parameter space of possible orbits. To better guide this parameter space exploration, we imposed the binary orbit to be close to coplanar with respect to the CBD and only mildly eccentric (below $0.2$; in agreement with \citealt{Hernandez+2024}). The latter choice is further motivated by the fact that, for unequal mass binaries, the overall disc morphology and the observed dust over-density in particular are well-reproduced for mildly eccentric and nearly coplanar binaries with respect to the CBD \citep{Poblete+2019}. Following this approach, we obtained the following possible binary orbital parameters: semi-major axis $a_{\rm b}=42.2$~au, eccentricity $e_{\rm b}=0.175$, inclination $i_{\rm b}=122.3\deg$, PA of ascending node $\Omega=168.9\deg$, argument of periastron $\omega=176.5\deg$, and initial true anomaly $f=163.5\deg$.

We use the 1-fluid method described in \cite{Laibe&Price2014} and \cite{Price&Laibe2015} to model a mixture of gas and dust using $10^6$ SPH particles, a sufficient  resolution to vertically resolve the disc. We set the grain size $s=0.1$~mm with intrinsic density $\rho_d=3\,{\rm g\,cm^{-3}}$ (typical of silicates). The dust size is considered to be constant and was selected to correspond to grains emitting at 1.3~mm since we are mainly interested in Band~6 ALMA observations. These particles are distributed between $R_{\rm in}=60$~au and $R_{\rm out}=140$~au, following an initial surface density profile defined as: $\Sigma=\Sigma_0 (R/R_0)^{-1}$, with $R_0=100$~au and $\Sigma_0=8.84\,{\rm g\,cm^{-2}}$. For completeness, we run an additional simulation where we set the dusty disc at $R_{\rm in}=80$~au (instead of 60 au) and find no significant differences compared to the nominal case (see Fig.~\ref{fig:comp6080}). Therefore, the initial condition does not strongly impact the emergence of the over-density. For the specific case of L1551 IRS~5, the surrounding material is divided into two components: the CBD and the infalling envelope, each with a different power-law index. Here we are mainly interested in the disc so we chose to neglect the envelope. The CBD aspect ratio if fixed to 0.05 at a distance of 100~au from the binary centre of mass, consistent with recent measurements for Class I discs \citep{Villenave+2023}. Once the simulations starts, the CBD feels the periodic gravitational perturbation of the inner binary (creating spirals, streamers, and the dust enhancement) and the disc also expands radially due to viscous forces \citep{phantom, Price+2018}.

We assume that initially the dust component is uniformly distributed within the gaseous disc and we adopt a typical value of $0.01$ for the dust-to-gas ratio. A simple estimate of the dimensionless Stokes number at the inner edge (60\,au) is:
\[
\mathrm{St}\simeq\frac{\pi}{2}\frac{\rho_{\rm s} s}{\Sigma_{\rm g}},
\]
using $\Sigma(60\,\mathrm{au})\simeq14.7\ \mathrm{g\,cm^{-2}}$, $s=0.1\,$mm and $\rho_{\rm s}=3\ \mathrm{g\,cm^{-3}}$ yields $\mathrm{St}\approx3\times10^{-3}$. Thus the 0.1\,mm particles in our runs are tightly coupled to the gas, fully validating the one–fluid algorithm adopted for the simulations. This strong coupling is also consistent with the behaviour seen in Fig.~1, where dust and gas follow nearly identical large-scale structures. We also note that considering icy aggregates with $\rho_{\rm s}=1\ \mathrm{g\,cm^{-3}}$ would not affect the coupling in a significant way (only by a factor of 3).

In addition, we estimate the Toomre parameter as follows:
\[
Q=\frac{c_s\Omega}{\pi G\Sigma}\simeq\frac{hM_\star}{\pi R^2\Sigma},
\]
using $c_s=h v_{\rm k}$ and $v_{\rm k}^2=GM_\star/R$. With $M_\star=1.1\,M_\odot$, $h\!=\!0.05$, and our surface-density profile ($\Sigma(60\,\mathrm{au})\simeq14.7\ \mathrm{g\,cm^{-2}}$, $\Sigma(100\,\mathrm{au})=8.84\ \mathrm{g\,cm^{-2}}$) we obtain $Q\approx3.0$ at 60\,au, $Q\approx1.8$ at 100\,au and $Q\approx1.3$ at 140\,au. These values (all $>1$) indicate that the circumbinary disc is globally stable against axisymmetric gravitational collapse under our assumed parameters. We note, however, that uncertainties in $\Sigma$, $h$ and any additional mass in the envelope could lower $Q$ locally; nevertheless, with the disc masses inferred for L1551 IRS\,5 ($\sim$0.02–0.03\,$M_\odot$) self-gravity is unlikely to dominate the dynamics in the region modelled.

The snapshot with the highest level of resemblance with the observations (across different wavelength and tracers) is are shown in the upper row of Figure~\ref{fig:panel-sims}.

\begin{figure}
    \centering
   \includegraphics[width=0.5\textwidth]{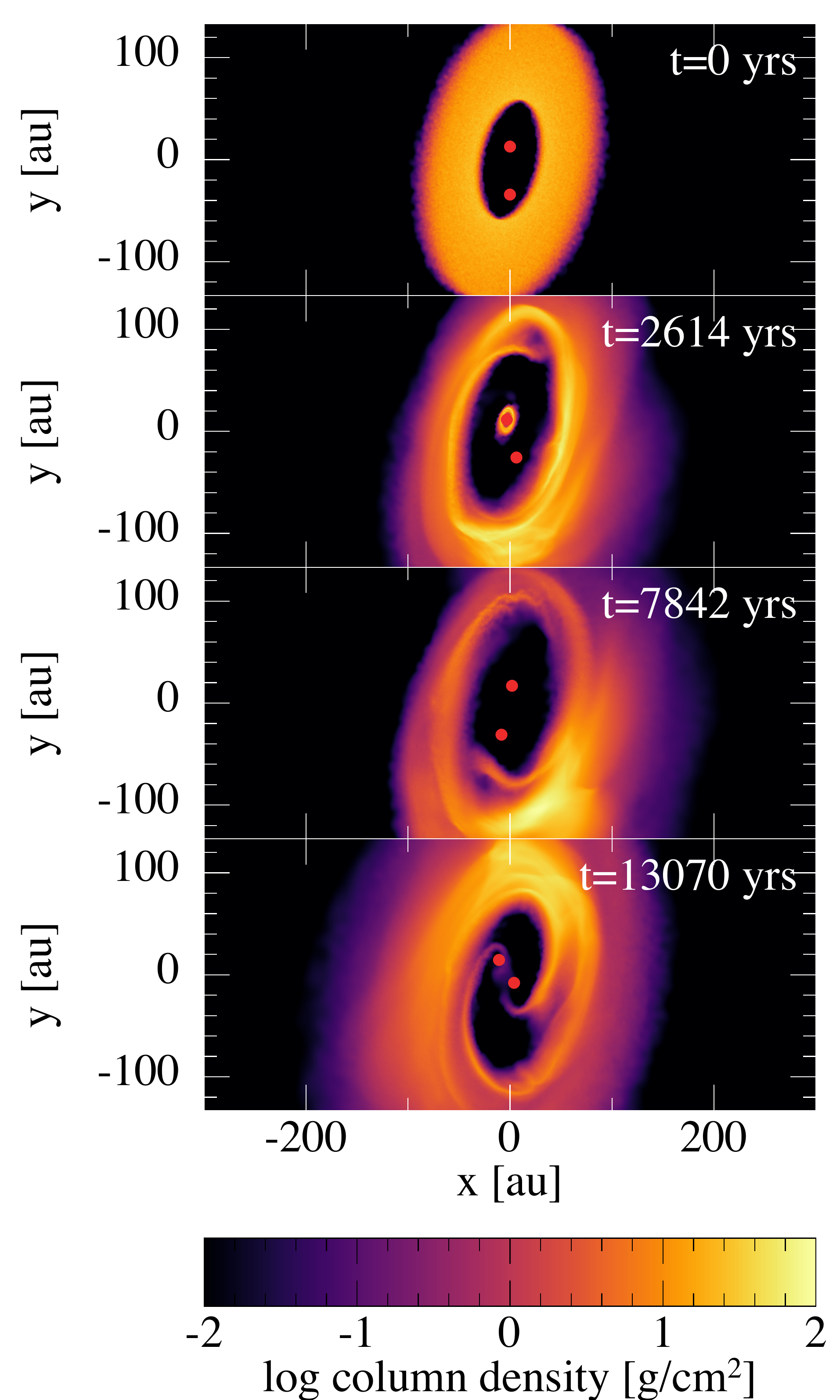}
    \caption{Gas surface density for different evolutionary stages for our nominal {\sc Phantom} hydrodynamical model. Time moves forward from top to bottom: 0, 10, 30, 50 binary orbits.}
    \label{fig:timeevol}
\end{figure}

\begin{figure}
    \centering
   \includegraphics[width=0.5\textwidth]{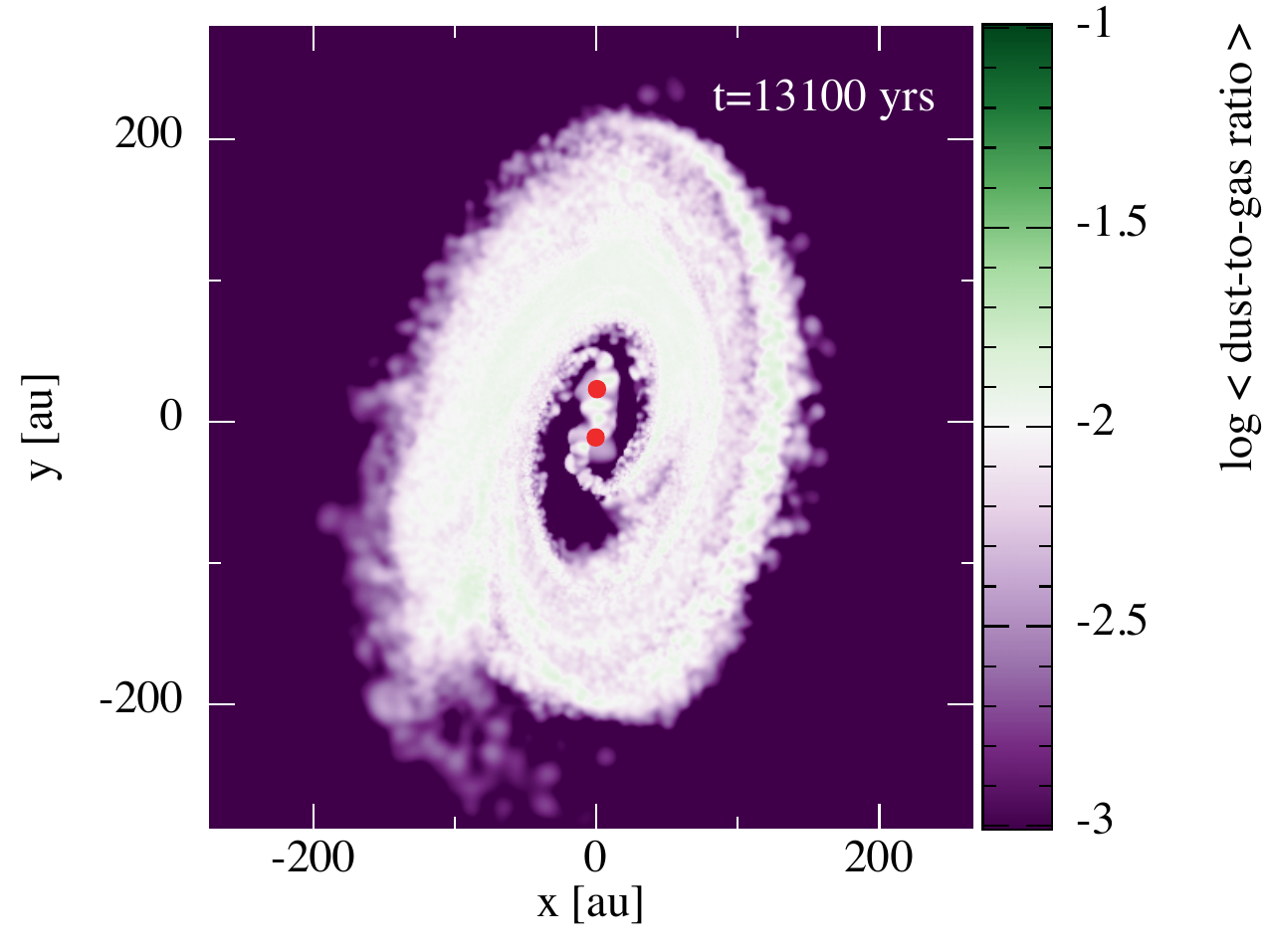}
    \caption{Dust-to-gas ratio after t=13\,100 years. The homogeneous distribution indicates that the dust over-density corresponds to a dust concentration rather than a dust trap for which the ratio should increase at the over-density location.}
    \label{fig:dgr}
\end{figure}

\begin{figure}
    \centering
   \includegraphics[width=0.5\textwidth]{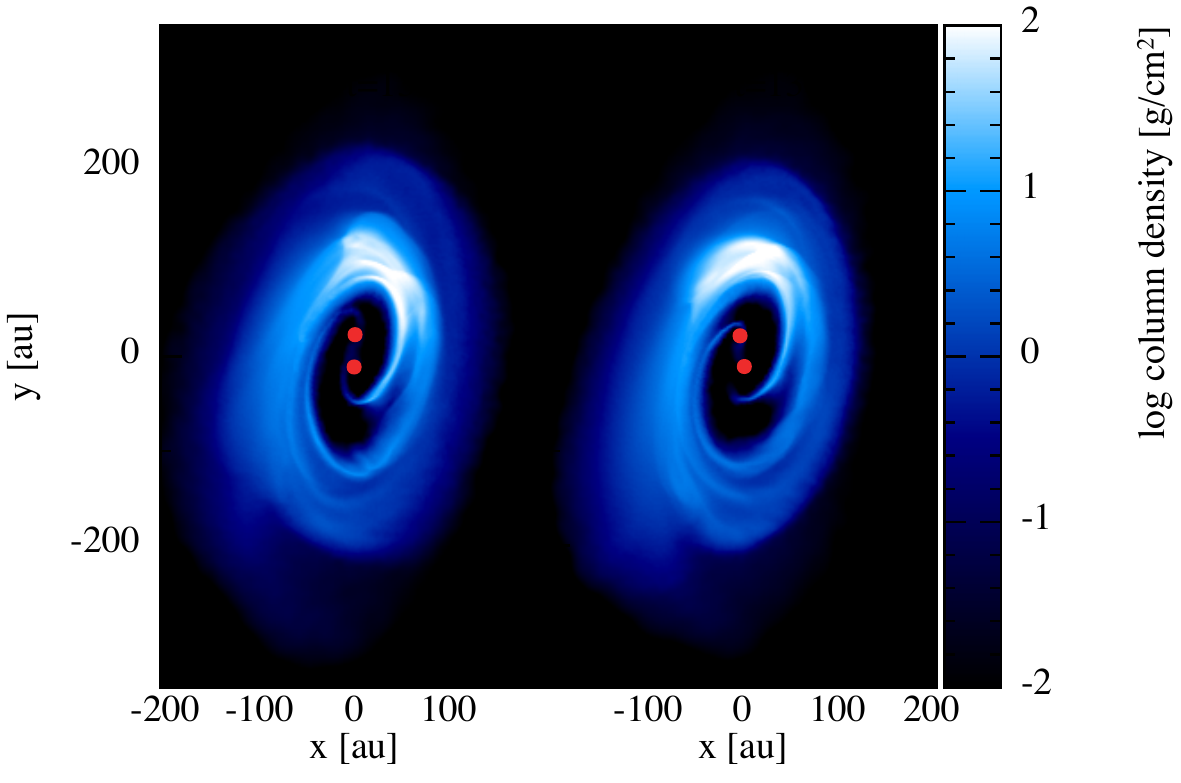}
    \caption{Comparison between two different initial conditions with $R_{\rm in}=60$~au (left panel) and $R_{\rm in}=80$~au (right panel) after t=13\,100 years. We do not see noticeable differences showing that the over-density is not a direct result of the initial condition.}
    \label{fig:comp6080}
\end{figure}

\subsection{Radiative transfer calculations}
\label{sec:mcfost}

In order to perform a direct comparison of the proposed model with the available observations of L1551 IRS 5, we post-processed a set of outputs from the hydrodynamical simulations using the public Monte Carlo radiative transfer code {\sc mcfost} (version 4.1, \citealt{Pinte+2006, Pinte+2009}). The Voronoi tessellation used in {\sc mcfost} is particularly well-suited to post-process data from {\sc Phantom} since each cell corresponds to the position of an SPH particle. Therefore, we map the density, temperature, velocity maps from the hydrodynamical simulations into the radiative transfer grid without any interpolation. 

The disc is irradiated using $1.28\times10^8$ photons packets, considering two sink particles of different luminosity as isotropic sources with stellar spectral models appropriate to their mass. In our calculations, we set the temperatures of the stars so the primary (N) and secondary (S) components have luminosities equal to $26.4\,L_{\odot}$ and $0.2\,L_{\odot}$ (respectively). The combined luminosity is in agreement with the available observations constraints \citep{Green+2013}. We assume astronomical silicates \citep{Draine2003}, with a grain size distribution ranging from 0.03~$\mu$m to~1 mm, and a slope of $-3.5$. Optical properties are calculated using Mie theory. The whole system (stellar binary and CBD) is observed from a distance of 141~pc and inclined using the same angles as in \cite{Bianchi+2020}.

Finally, we predict line emission from the gas content of the disc for the isotopologues of carbon monoxide C$^{18}$O. Furthermore, we suppose that the gas is in local thermodynamic equilibrium with $T_{\rm gas} = T_{\rm dust}$. Last, we compute the thermal emission from dust in the continuum based on the dust distribution in the disc. With these emission maps and data cubes, we then produce the synthetic images presented in the bottom row of Figure~\ref{fig:panel-sims}.

\end{appendix}

\end{document}